\documentclass[aps,prl,showpacs,twocolumn]{revtex4}
\usepackage{amssymb}

\usepackage{graphicx}

\begin{document}

\title{Topologically distinct critical theories emerging from the bulk
entanglement spectrum of integer quantum Hall states on a lattice}
\author{Qiong Zhu$^{1}$, Xin Wan$^{1}$, and Guang-Ming Zhang$^{2,3}$}
\affiliation{$^{1}$Zhejiang Institute of Modern Physics, Zhejiang University, Hangzhou
310027, China\\
$^{2}$State Key Laboratory of Low-Dimensional Quantum Physics and Department
of Physics, Tsinghua University, Beijing 100084, China\\
$^{3}$Collaborative Innovation Center of Quantum Matter, Beijing, China}
\date{\today}

\begin{abstract}
The critical theories for the topological phase transitions of integer
quantum Hall states to a trivial insulating state with the same symmetry can
be obtained by calculating the ground state entanglement spectrum under a
symmetric checkerboard bipartition. In contrast to the gapless edge
excitations under the left-right bipartition, a quantum network with bulk
gapless excitations naturally emerges without fine tuning. On a large finite
lattice, the resulting critical theory for the $\nu =1$ state is the (2+1)
dimensional relativistic quantum field theory characterized by a \textit{%
single} Dirac cone spectrum and a pair of \textit{fractionalized}
zero-energy states, while for the $\nu =2$ state the critical theory
exhibits a parabolic spectrum and no sign of fractionalization in the
zero-energy states. A triangular correspondence is established among the
bulk topological theory, gapless edge theory, and the critical theory via
the ground state entanglement spectrum.
\end{abstract}

\maketitle

\textit{Introduction.-} The integer quantum Hall (IQH) effect has been well
studied in various two-dimensional (2D) electron systems, including the
experimental observation in graphene\cite{novoselov05,zhang05}. When the
time reversal symmetry is broken, the 2D electron systems are characterized
by a topological quantum number known as the Chern number\cite{thouless82},
which can be measured experimentally as the quantized Hall conductance with
high precision\cite{vonklitzing80}. In 1988, Haldane\cite{haldane88}
proposed a famous honeycomb lattice model for the IQH effect, which allows
the access to the phase transition between the pure Hall state ($\nu =1$)
and its trivial insulating state ($\nu =0$) through tuning the time-reversal
symmetry or inversion symmetry breaking parameters. Such a transition is a
prototype of topological quantum phase transition. Unlike the
Landau-Ginzberg-Wilson paradigm for the symmetry breaking phase transitions,
the quantum criticality is characterized by the Dirac spectrum with the
parity anomaly and undoubled chiral fermions in the lattice realization of
the (2+1)D relativistic quantum field theory\cite{haldane88,ludwig94}.
Moreover, interesting features can arise in the phase transitions for the $%
\nu >1$ IQH states, e.g., charge carriers with the parabolic energy spectrum
in pristine bilayer graphene exhibit distinct features with a double-step
transition\cite{novoselov-geim06}. Therefore, a systematic method is called
for to study the critical properties of the 2D topological phases.

It is generally believed that the essential information of a topological
phase is encoded in its ground state wave function. Li and Haldane\cite{li08}
introduced entanglement spectrum as the collection of eigenvalues of the
reduced density matrix under a left-right bipartition, and found that the
low-lying spectrum bears a remarkable similarity to the physical edge
spectrum of the topological state\cite{thomale10,chandran11,qi12,alba12}. A
further step is to ask how one can extract the bulk critical properties for
the transition from the topological phase to its trivial insulating phase
with the same symmetry, and what the relationship is between the
corresponding gapless edge theory and the bulk critical theory.

In fact, the critical theory can be regarded as the ``domain wall'' between
the topological phase and its trivial insulating phase in the parameter
space, hence realizing delocalized boundary excitations of the non-trivial
topological phase\cite{chen13}. To reveal such a non-trivial quantum
criticality, appropriated boundary conditions are needed. For the IQH
transitions, the domains of trivial and non-trivial phases described by a
Dirac Hamiltonian with respective positive and negative masses\cite%
{haldane88,ludwig94} must be adjacent to each other in the model for the
critical theory. Accordingly, the edge modes along the boundaries of the
domains form a percolating network. As Haldane pointed out\cite{haldane88},
along the critical line between the trivial and topological phases, one
expects a low-lying massless Dirac spectrum. Hsieh and Fu\cite{hsieh13}
argued that a symmetric extended bipartition of the system into two
subsystems as long as the size of the sublattice unit cell is larger than
the correlation length of the topological phase can realize such boundary
conditions.

According to Ref.~\cite{hsieh13}, the bulk entanglement spectrum of the IQH
state with $\nu =1$ (or odd integers) is gapless, while that of a state
characterized by an even Chern number may be gapped. This is
counterintuitive because the conclusion suggests that the bulk entanglement
spectrum of a $\mathbb{Z}$ topological insulator is characterized by a $%
\mathbb{Z}_{2}$ index; in other words, the bulk entanglement spectrum and
the edge theory do not match. Should there, then, be a one-to-one
correspondence at all between the edge structure of a topological phase and
its critical bulk entanglement spectrum? Does the emergent symmetries at the
critical point play a role for the mismatch? In this Letter, we answer these
questions by analyzing the bulk entanglement spectrum in a lattice model of
IQH states with $\nu =1$ and, more importantly, $\nu >1$. We show that the
critical theory for the $\nu =1$ state is the (2+1)-dimensional relativistic
quantum field theory characterized by a single Dirac cone spectrum and a
pair of fractionalized zero-energy states. For the $\nu =2$ state, the bulk
entanglement spectrum exhibits a quadratic band crossing. The spectral
difference resembles vividly that between monolayer graphene\cite%
{novoselov05,zhang05,castroneto09} and pristine bilayer graphene\cite%
{novoselov-geim06,mccann06}. In general, we find that the Chern index of the
topological phase is in one-to-one correspondence with the Berry flux for
the ground state wave function of the entanglement Hamiltonian. Extending
the bulk-edge correspondence of a topological phase, we propose a generic
triangular correspondence among the bulk topological theory, gapless edge
excitations, and the critical theory as revealed by the bulk entanglement
spectrum.

\begin{figure}[tbp]
\includegraphics[width=8cm]{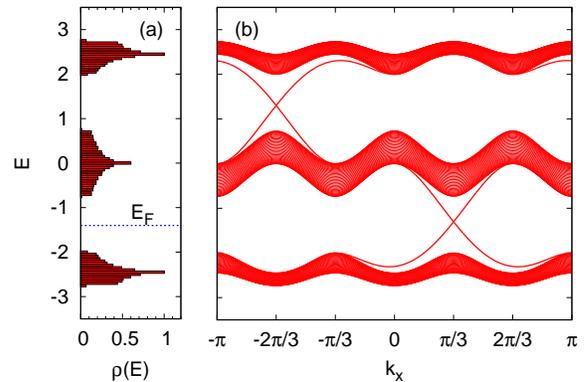}
\caption{(Color online) (a) Density of states of the tight-binding $%
300\times 300$ lattice model with periodic boundary conditions. (b)
Corresponding dispersion curves with open boundary condition in $y$%
-direction only. }
\label{fig:dos}
\end{figure}

\textit{Model and method.-} We consider a tight-binding Hamiltonian on a 2D
square lattice in the presence of a perpendicular magnetic field,
\begin{equation}
H=-\sum_{\langle ij\rangle }{\left( e^{i\theta _{ij}}c_{i}^{\dagger
}c_{j}+h.c.\right) },  \label{eq:hamiltonian}
\end{equation}
where the total phase that an electron picks up when moving around a
plaquette
\begin{equation}
\sum_{\Box }\theta _{ij}=2\pi \frac{\phi }{\phi _{0}}
\end{equation}
is given by the magnetic flux $\phi $ per unit cell, in units of magnetic
flux quantum $\phi _{0}=hc/e$. The energy spectrum of the system exhibits a
self-similar pattern, known as the Hofstadter butterfly\cite{hofstadter76}.
We first focus on the case of $\phi /\phi _{0}=1/3$, at which the
tight-binding band is split into three magnetic subbands, which carry Chern
numbers $\nu =1$, $-2$, and $1$, respectively. Figure~\ref{fig:dos}(a) shows
the density of states from a $300\times 300$ square lattice. With the open
boundary condition in the $y$ direction, the corresponding spectrum includes
chiral edge states in the two band gaps, as illustrated in Fig.~\ref{fig:dos}%
(b). Actually the tight-binding Hamiltonian can describe a lattice
realization of the Landau levels and the subsequent IQH transitions. The
topological phases belong to the class A in the classification of
noninteracting topological insulators/superconductors\cite{schnyder08}. The
time-reversal symmetry is explicitly broken by the external magnetic flux,
which determines the propagation direction of the chiral edge modes.

The presence of the chiral edge modes in the topological system is an
example of the so-called bulk-edge correspondence. It also manifests at the
artificial boundary in the study of quantum entanglement\cite{li08}. After
we fill up the lowest magnetic subband in the case of $\phi =\phi _{0}/3$,
the Fermi energy lies in the spectral gap and the system is in the $\nu =1$
IQH phase. If we cut a square A in the bulk of the system and trace out all
the degrees of freedom in area B outside the square in the density matrix of
the ground state $\Psi _{0}$, the spectrum of the resulting reduced density
matrix $\rho _{A}=$Tr$_{B}(|\Psi _{0}\rangle \langle \Psi _{0}|)$ resembles
the energy spectrum of a chiral Fermi liquid, which is described by a (1+1)D
chiral massless Dirac fermion field theory. Heuristically, we can represent
such a bipartition by a directed loop surrounding A.

The technique on computing the spectrum of the reduced density matrix for
non-interacting systems has been well documented in the literature\cite%
{peschel09}. The key lies in the computation of the correlation matrix $%
g_{ij}=\langle c_{i}^{+}c_{j}\rangle $ in the ground state, where $i$, $j$
denote the lattice sites in the subsystem A. The correlation matrix is
isospectral to the reduced density matrix $\rho _{A}\equiv e^{-H_{E}}$ or
the entanglement Hamiltonian $H_{E}$. For a non-interacting electron system,
the entanglement Hamiltonian takes a quadratic form
\begin{equation}
H_{E}=\sum_{n,m}h_{nm}c_{n}^{+}c_{m}.
\end{equation}
More precisely, $g_{ij}$ encodes the Fermi-Dirac distribution function for
the eigenstates of $H_{E}$ such that the eigenvalues of $g_{ij}$ are
expressed as $\eta _{l}=1/(e^{\varepsilon _{l}}+1)$, in which $\varepsilon
_{l}$s are the eigenvalues of $H_{E}$, or the energy levels of the ground
state entanglement spectrum. As usual, the states with negative $\varepsilon
_{l}$ are occupied.

\begin{figure}[tbp]
\includegraphics[width=8cm]{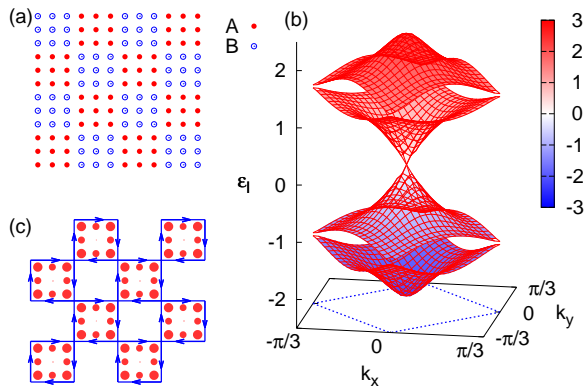}
\caption{(Color online) (a) Symmetric checkerboard bipartition illustrated
in a $12\times 12$ lattice. (b) Spectrum of the entanglement Hamiltonian $%
H_{E}$. For the clarity of the Dirac spectrum centered at $k=(0,0)$ and $%
\protect\varepsilon _{l}=0$, we only plot the four bands with energies
closest to zero. The Brillouin zone corresponding to the checkerboard
lattice is illustrated by the blue square. (c) The total probability density
of the two zero-energy states in the finite $12\times 12$ lattice, whose
chiral edge-state nature is represented by the directed loops. The area of
each dot is proportional to the local probability density. }
\label{fig:cone}
\end{figure}

\textit{Emergence of a single Dirac cone.-} To reveal the information beyond
the edge properties of the IQH $\nu =1$ state, we implement the symmetric
checkerboard bipartition of a finite lattice\cite{hsieh13}, as illustrated
in Fig.~\ref{fig:cone}(a). Heuristically, the remaining half of the square
blocks supports chiral edge modes (one per block) represented by directed
loops, which form a regular lattice with the unit cell twice the size of the
block and rotated by $\pi /4$. With quantum tunneling at the block corners,
the loops coalesce into a regular network. If we further assign random
tunneling amplitudes at the corners, we will obtain the Chalker-Cottington
network model\cite{chalker88}, which can describe the generic IQH plateau
transition.

Figure~\ref{fig:cone}(b) plots the spectrum of the entanglement Hamiltonian $%
H_{E}$ for a large square lattice with a linear size $L=300$. Here, we
choose each block in part A or B to include $3\times 3$ sites, or a total
magnetic flux $3\phi _{0}$. For clarity, we only plot the four bands (out of
nine) closest to zero in the entanglement spectrum. The most remarkable
feature of the spectrum is the single Dirac cone appearing at the Brillouin
zone center. The emergent entanglement spectrum hence differs qualitatively
from the energy spectrum of the original lattice model. For comparison, we
note that the Hamiltonian~(\ref{eq:hamiltonian}) with $\phi =\phi _{0}/2$
has two Dirac cones in its energy spectrum that usually appears in lattice
realizations of the (2+1)D relativistic quantum field theories. The defeat
of the fermion doubling here is rooted both in the broken time-reversal
symmetry of the original lattice model and in the broken 2D parity symmetry
due to the checkerboard bipartition. Accordingly, the bipartition generates
a non-interacting Hamiltonian $H_{E}$ on the non-simply connected 2D
sublattice A with long-range ($\sim 1/r$) complex hopping amplitudes.

At the Dirac cone vertex, there \textit{always} exist two degenerate
zero-energy states, regardless of the size of the lattice system
and the block size for the checkerboard bipartition; the Kramers degeneracy
is the consequence of the emergent time-reversal symmetry of the low-energy
effective Dirac Hamiltonian $H_1 = \alpha \mathbf{\sigma} \cdot \mathbf{p}$%
\cite{ludwig94}. We plot the \textit{total} probability density of the
zero-energy states in Fig.~\ref{fig:cone}(c) by filled circles. In each $%
3\times 3$ block the density retains the point-group symmetry of the lattice
and the center site in each block has zero density. This is a striking
manifestation of the constituting edge mode within each block. With the
directed loops attached around the blocks, we emphasize the emergence of the
disorder-free Chalker-Cottington network picture\cite{chalker88}, which can
be mapped to the Dirac Hamiltonian in 2D~system\cite{ho96}. In fact, the
eigenstates of the whole Dirac cone are linear combinations of the edge
modes of individual blocks.

\begin{figure}[tbp]
\includegraphics[width=8cm]{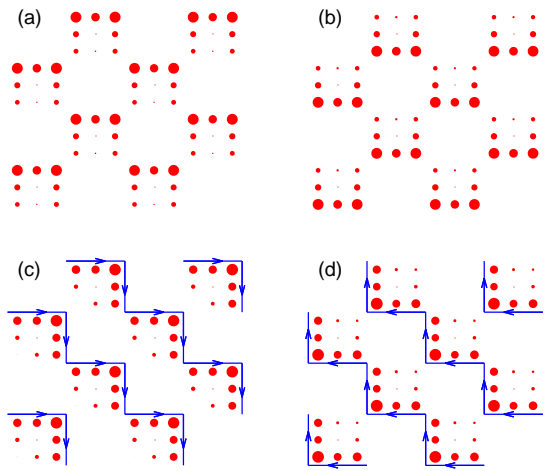}
\caption{(Color online) (a) and (b) The probability densities (proportional
to the area of the dots) of the two zero-energy states (we only show a $%
12\times 12$ sector for clarity). The probability is significant on half of
the edge sites from each block only. (c) and (d) The probability densities
of the two zero-energy states in (a) and (b) after a rotation of $\protect%
\pi /8$ in their spinor representation. The sites with significant
probability density follow chiral stairs either up or down. }
\label{fig:zeroModes}
\end{figure}

\textit{Fractionalization of zero-energy states.-} The single Dirac cone
structure in the entanglement spectrum also demonstrates that the low-energy
effective Hamiltonian $H_1$ has a particle-hole symmetry, which is not
present in the original lattice model\cite{haldane88,ludwig94}. A further
consequence is the charge fractionalization due to the parity anomaly of the
(2+1)D effective field theory\cite{jackiw84}. We have carefully studied this
two-dimensional space spanned by the zero-energy states at the Dirac cone
vertex by plotting their separate probability densities. As illustrated in
Fig.~\ref{fig:zeroModes}(a) and (b), the fractionalization manifests in a
way that either state is formed by the significant occupations on only
\textit{half} of the edge sites in each block. The fractionalization cannot
be removed by any linear combinations of the two states, which only rotate
the probability density around the edge in each block simultaneously.

The fractionalization of the edge degrees of freedom of each block is
consistent with the fact that the Dirac cone emerges at the center of the
Brillouin zone, hence representing the long-wavelength collective behavior
of the system. By a $\pi /8$-rotation of the zero-energy states in their
spinor representation, the probability densities rotate $\pi /4$ in the real
space along the edge of each block, and we effectively fractionalize the
network model as in Fig.~\ref{fig:zeroModes}(c) and (d). One state has the
probability density essentially on the edges marked by right and down arrows
- the right mover, while the other by left and up ones - the left mover.
They form chiral, coherent, and penetrating paths across the system along
the diagonal direction. The features of the fractionalization become more
prominent as we increase the block size in each sublattice, which allows a
cleaner separation of the bulk and the edge of each block. According to the
Jackiw argument\cite{jackiw84}, the charge conjugation symmetry leads to a
quantum Hall effect at $\nu =1/2$ with the zero-energy states filled and at $%
\nu =-1/2$ with them empty.

\textit{Topologically distinct transitions.-} The low-lying massless
spectrum of the undoubled chiral fermions and the parity anomaly confirms
that the phase transition from the $\sigma _{xy}=e^{2}/h$ IQH state to an
insulator can emerge from the entanglement spectrum of the ground state wave
function in the non-trivial phase. In unspecific words, we can unveil the
full information of the gapless critical point from the wave function on the
topological side. To support the general statement, we continue to explore
the emergent physics under the checkerboard bipartition for the IQH state
with $\sigma _{xy}=2e^{2}/h$.

\begin{figure}[tbp]
\includegraphics[width=8cm]{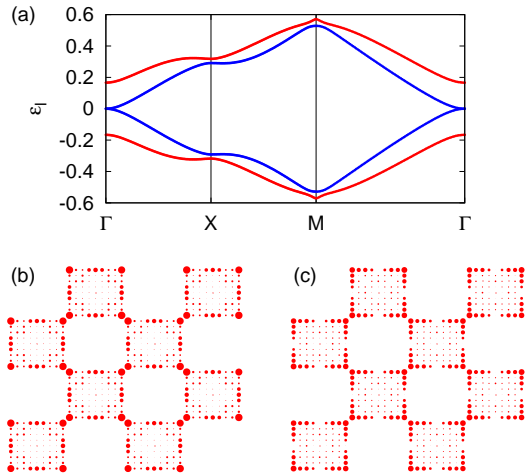}
\caption{(Color online) (a) Dispersion curves of the entanglement spectrum
from the checkerboard bipartition (in $9\times 9$ blocks) with $\protect\phi %
= \protect\phi _{0}/9$ and two magnetic subbands filled. (b) and (c) The
probability densities (proportional to the area of the dots) of the two
zero-energy states. }
\label{fig:twoBands}
\end{figure}

In Figure~\ref{fig:twoBands}(a), we plot the entanglement spectrum in a $%
\phi =\phi _{0}/9$ system with the lowest two magnetic subbands filled,
which is a topological state with $\sigma _{xy}=2e^{2}/h$. In contrast to
the Dirac cone spectrum in the $\sigma _{xy}=e^{2}/h$ case, we find 
undoubled chiral fermions with a parabolic dispersion associated with a
Berry flux of $2\pi$. In addition, a pair of adjacent bands [red lines in
Fig.~\ref{fig:twoBands}(a)] is also visible with a gap opened at the
Brillouin center. The band structure, together with the earlier $\sigma
_{xy}=e^{2}/h$ case, reminds us of the spectral difference between monolayer
and pristine bilayer graphene\cite{castroneto09}, except that the graphene
systems have two valleys at the two inequivalent corners of their hexagonal
Brillouin zone. In our case, the parabolic low-energy spectrum emerges at
the zone center without fine tuning. The four bands can be regarded as
resulting from the reconstruction of the two Dirac cones which one may
naively expect to appear at the zone center, if one asserts that no quantum
entanglement exists between states in different subbands separated by a
spectral gap. The existence of the quadratic band-crossing point is allowed
due to the emergent time-reversal symmetry and the $C_4$ lattice symmetry in
the original model\cite{sun09}. We further find that, like graphene, the
distinction of the $\sigma _{xy}=e^{2}/h$ and $2e^{2}/h$ cases lies in the
Berry flux ($\pi $ and $2\pi $, respectively) of the chiral fermions\cite%
{mccann06}. Therefore, the emergent spectrum describes a double-step
transition from the $\sigma _{xy}=2e^{2}/h$ IQH state to the corresponding $%
\sigma _{xy}=0$ state, much like the double-step IQH transition in bilayer
graphene\cite{novoselov-geim06}.

The fractionalization of the zero-energy states is the hallmark of the
(2+1)D relativistic quantum field theory, hence ought to be absent in the $%
\sigma _{xy}=2e^{2}/h$ case. We plot the probability densities of the two
zero-energy states in Fig.~\ref{fig:twoBands}(b) and (c). As was pointed out
earlier, the densities are mostly distributed along the block edges.
However, we find no sign of fractionalization as in the earlier case, even
though we exhaust all linear combinations of the states, which only lead to
the probability density transfer between the corners and the edges (somewhat
like breathing).

We further study the IQH states with $\nu >2$. For example, four Dirac
points with Berry flux $\pi$ and one additional Dirac point with flux $-\pi$
can appear for $\nu = 3$, and four Dirac points with Berry flux $\pi$ can
appear for $\nu = 4$, which are consistent with the $C_4$ lattice symmetry
in the Hoftadter lattice model. In general, the number of the Dirac cones
can be affected by the reconstruction of the entanglement spectrum, but the
associated Berry flux cannot. In other words, the Berry flux for the ground
state wave function of the entanglement Hamiltonian provides a topological $%
\mathbb{Z}$ classification of the transitions from the original IQH states
to the trivial one.

\textit{Conclusion and discussion.-} We have shown that the critical theory
between the topological state and the corresponding trivial state can emerge
in the entanglement spectrum of the non-trivial state with a symmetric
checkerboard bipartition. The bipartition generates a quantum network with
gapless bulk excitations. The critical theories for topologically different
systems can be distinguished by the behavior of the low-energy entanglement
spectrum and its degenerate zero-energy states. The IQH states to
insulator transitions we discuss here are in the clean limit, which may be
tuned by an additional periodic potential\cite{wen93}. Distinct effects of
quenched disorder may be considered separately.

Our results suggest that the Chern index characterizing the IQH states
corresponds faithfully to the Berry flux of the ground state wave function
of the entanglement Hamiltonian. The emergent particle-hole symmetry in the
entanglement spectrum \emph{does not change} the characterization of the
critical theory, even though it manifests in the degenerate zero-energy
states in the entanglement Hamiltonian. The bulk entanglement spectra for
both $\nu =1$ and $\nu =2$ IQH states are gapless. This contrasts to the
claim in Ref.~\cite{hsieh13} and the difference may be attributed to the
additional symmetries in their model Hamiltonian unnecessary for the IQH
states.

Remarkably, the ground state wave function of the topological bulk theory
encodes not only the boundary theory\cite{li08}, but also the critical
theory for its transition to a trivial insulator with the same symmetry,
both of which can be derived from the ground state entanglement spectrum
under different types of bipartitions\cite{hsieh13,rao14,hsieh14,santos14}.
Therefore, we can expect a generic triangular correspondence among the bulk
topological theories, the gapless edge theories, and the bulk critical
theories. The edge theories is the critical theories spatially confined
between the topological phases and the trivial insulating phases, while the
percolating quantum networks of the edge modes establish the critical
theories for the transitions of the topological phases to the trivial phases.

One of the authors (XW) would like to thank M.-C. Chung for many valuable
discussions on the entanglement spectrum in free lattice models. The authors
acknowledge the support of the 973 Program under Project No. 2012CB927404
and of NSF-China through the grants No. 20121302227 and No. 11174246.

\end{document}